\documentclass[10pt,preprint]{aastex}

\slugcomment{To appear in MNRAS Main Journal.}

\shorttitle{On the common origin of the AB Dor moving group and
the Pleiades cluster} \shortauthors{V. Ortega et al.}

\begin{document}


\title{On the common origin of the AB Dor moving group and the Pleiades cluster}

\author{V. G. Ortega\altaffilmark{1}, E. Jilinski \altaffilmark{1, 2}, R. de la Reza\altaffilmark{1} and B. Bazzanella\altaffilmark{1}}

\altaffiltext{1}{Observat\'orio Nacional, Rua General Jos\'e
Cristino 77, S\~{a}o Cristov\~{a}o, 20921-400, Rio de Janeiro,
Brazil.}

\altaffiltext{2}{Pulkovo Observatory, Russian Academy of Science,
65, Pulkovo, 196140 St. Petersburg, Russia.}

\email{vladimir@on.br, jilinski@on.br, delareza@on.br,
bruno@on.br}

\begin{abstract}
AB Dor is the nearest identified moving group. As with other such
groups, the age is important for understanding of several key
questions. It is important, for example, in establishing the
origin of the group and also in comparative studies of the
properties of planetary systems, eventually surrounding some of
the AB Dor group members, with those existing in other groups. For
AB Dor two rather different estimates for its age have been
proposed: a first one, of the order of 50 Myr, by Zuckerman and
coworkers from a comparison with Tucana / Horologium moving group
and a second one of about 100$-$125 Myr by Luhman and coworkers
from color-magnitude diagrams (CMD). Using this last value and the
closeness in velocity space of AB Dor and the Pleiades galactic
cluster, Luhman and coworkers suggested coevality for these
systems. Because strictly speaking such a closeness does not still
guarantee coevality, here we address this problem by computing and
comparing the full 3D orbits of AB Dor, Pleiades, $\alpha$ Persei
and IC 2602. The latter two open clusters have estimated ages of
about $85-90$ Myr and 50 Myr. The resulting age 119 $\pm$ 20 Myr
is consistent with AB Dor and Pleiades being coeval. Our solution
and the scenario of open cluster formation proposed by Kroupa and
collaborators suggest that the AB Dor moving group may be
identified with the expanding subpopulation (Group I) present in
this scenario. We also discuss other related aspects as iron and
lithium abundances, eventual stellar mass segregation during the
formation of the systems and possible fraction of debris discs in
AB Dor group.
\end{abstract}


\section{Introduction}

In a series of publications Eggen introduced and discussed the
concept of superclusters defined as stellar aggregates in the
solar neighborhood which are characterized by parallel space
motions of their members. One of these superclusters is the Local
Association (Eggen 1975, 1983 a,b) located within a radius of a
few hundred parsecs around the Sun and often referred to as the
Pleiades supercluster because of the similarity of the space
motions of its members and the Pleiades open cluster. Apart from
Pleiades, other open star clusters as $\alpha$ Persei, IC 2602,
$\delta$ Lyrae, NGC 1039, NGC 2516 and also the Scorpio-Centaurus
(Sco-Cen) OB association would be contained in the Local
Association.

With recent identifications of relatively small nearby groups or
associations of post T-Tauri stars with ages from 8 Myr to 50 Myr
(Zuckerman and Song 2004b, Torres et al. 2006), located within
100$-$150 pc of the Sun and with previous knowledge of the Sco-Cen
complex (de Zeeuw et al. 1999, Mamajek et al. 2002), it can be
said that a new stage of investigation of the structure of the
Local Association has begun. We are now even in a position to
suggest plausible formation scenarios for some of those groups.
For instance, the young kinematic groups TW Hya (Torres et al.
2003, Zuckerman \& Song 2004b) with an age of 8.3 Myr (de la Reza
et al. 2006), $\beta$ Pictoris (Zuckerman et al. 2001, Torres et
al. 2006) with an age of 11.3 Myr (Ortega et al. 2002, 2004) and
$\eta$ and $\epsilon$ Chamaeleontis groups (Mamajek et al. 2000,
Feigelson et al. 2003) with an age of 6.7 Myr (Jilinski et al.
2005) may be the result of a sequence of bursts of low mass star
formation related to the older Sco-Cen subgroups with an age of
about 18 Myr (Ortega et al. 2005).

In addition to these, two other groups have been identified within
50 pc of the Sun: the Tucana / Horologium (Zuckerman et al. 2000,
Torres et al. 2000) and AB Dor (Zuckerman et al. 2004a) moving
groups. For the AB Dor group, comoving with the star AB Doradus
only 15 pc from the Sun, Zuckerman et al. (2004a) derived an age
of about 50 Myr on the basis of a comparison with the Tucana /
Horologium group (Zuckerman and Webb 2000, Torres et al. 2000,
Zuckerman and Song 2004b) and from the location of AB Dor M
spectral type stars in the CMD. However, Luhman et al. (2005)
analyzing the discrepancy between evolution models for very low
mass stars and direct mass measurements of the brown dwarf C
component of the triple AB Doradus system (Close et el. 2005,
Nielsen et al. 2005, Marois et al. 2005), have proposed that the
AB Dor moving group may have larger age in the 100$-$125 Myr
range. Moreover, by noticing that the mean space motion of AB Dor
group is similar to the Pleiades galactic cluster's mean motion,
Luhman et al. (2005) suggested that the former is probably coeval
with the latter. In this context, it is interesting that Innis et
al. (1986) had proposed that the rapid rotators AB Doradus and PZ
Tel may be part of the group of fast rotating K stars of the
Pleiades cluster, what would characterize both of them as young
systems. As for PZ Tel, we now know that this star is a member of
the $\beta$ Pictoris moving group (Zuckerman et al. 2001, Torres
et al. 2006) for which we have a fairly accurate age estimate. In
contrast, for AB Dor we have two conflicting age values.

To resolve this discrepancy and derive an age estimate for the AB
Dor group here we tackle the problem by calculating the Galactic
orbits of both the Pleiades cluster and the AB Dor group. We also
compare the orbits of AB Dor with those of the $\alpha$
\textbf{Persei} and IC 2602 clusters, which are considered to be
younger than Pleiades (Makarov 2006, Randich 2001b, Randich et al.
2001a) with estimated ages of $\sim$ 50 Myr comparable to the age
of AB Dor moving group as given by Zuckerman et al. (2004a). For
the $\alpha$ Persei cluster Stauffer et al. (1999) and Barrado y
Navascu\'es et al. (2004)) have estimated ages of 90 $\pm$ 10 Myr
and 85 $\pm$ 20 Myr, respectively, on the basis of lithium
depletion.

\section{The orbits }

As mentioned above, one of the main reasons given by Luhman et al.
(2005) for coevality of AB Dor with the Pleiades cluster was the
similar space motions of these groups. Nevertheless, this
similarity is not a sufficient condition for coevality. The full
3D path for each system has to be calculated.

As in previous work (e.g. Ortega et al. 2002), our method consists
in integrating back in time the 3D orbits of the stars and
following the evolution to check for the existence of a first
minimum of the configuration. When such a minimum exists the
orbits are confined in a certain region which is considered their
place of birth. Equivalently, when a comparison of the orbits of
the centers of two stellar systems is made, the minimum distance
gives the radius of the region occupied by those systems. The
dynamical age is then determined by the time elapsed from orbital
convergence. The resulting age is called $\it dynamical$ because
the orbits are calculated under a (modelled) Galactic potential.

\begin{figure}
\epsscale{0.7} \plotone{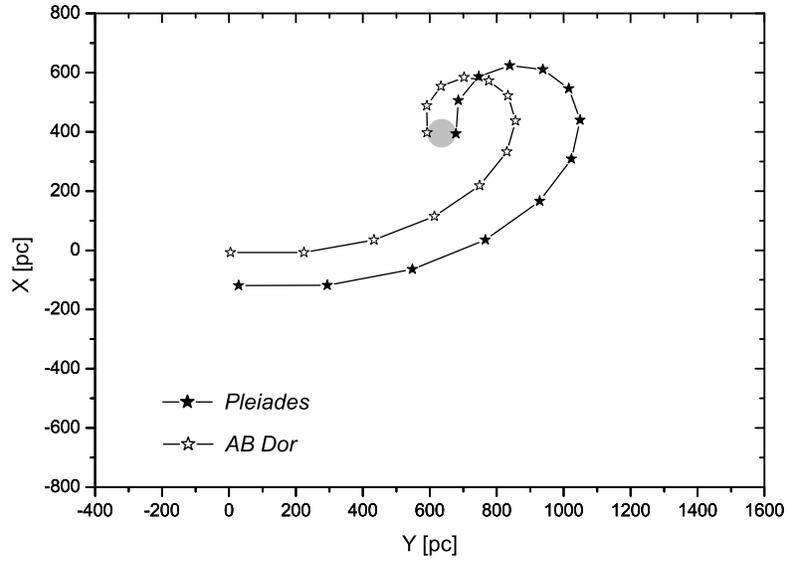} \caption{Galactic orbits of
Pleiades cluster and AB Dor moving group in projection on the XY
plane. The size of the grey circle corresponds to the size of the
birth region (R$=$43 pc). Each two point interval corresponds to
10 Myr time interval. \label{fig1}}
\end{figure}

\begin{figure}
\epsscale{0.7} \plotone{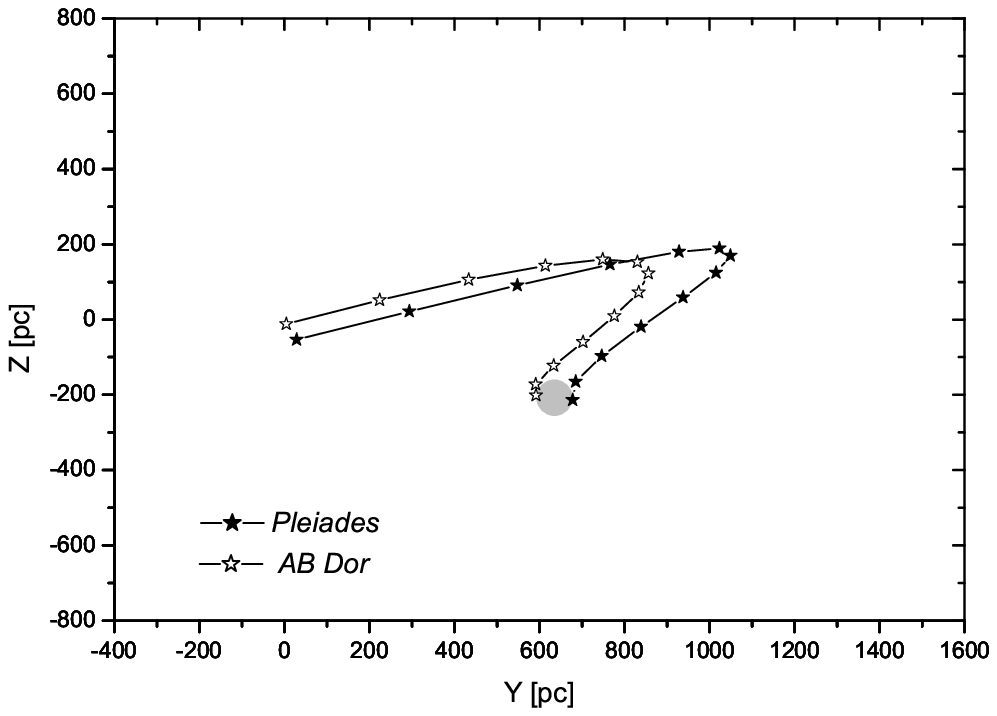} \caption{The same as in Fig 1 in
projection on the YZ Plane. \label{fig2}}
\end{figure}

\begin{figure}
\epsscale{0.7} \plotone{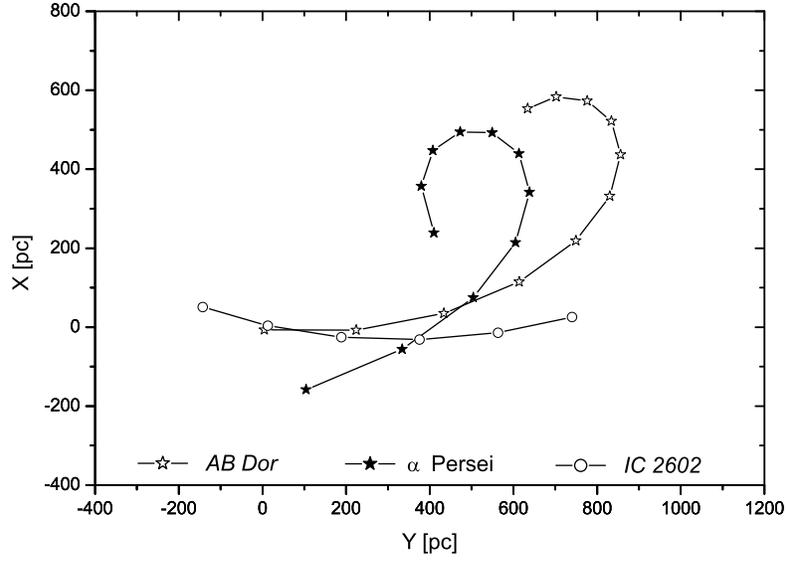} \caption{Galactic orbits of AB
Dor moving group and $\alpha$ Persei up to $-$ 100 Myr and IC 2602
clusters up to $-$50 Myr in projection on the XY plane. Each two
point interval corresponds to 10 Myr time interval \label{fig3}}
\end{figure}

\begin{figure}
\epsscale{0.7} \plotone{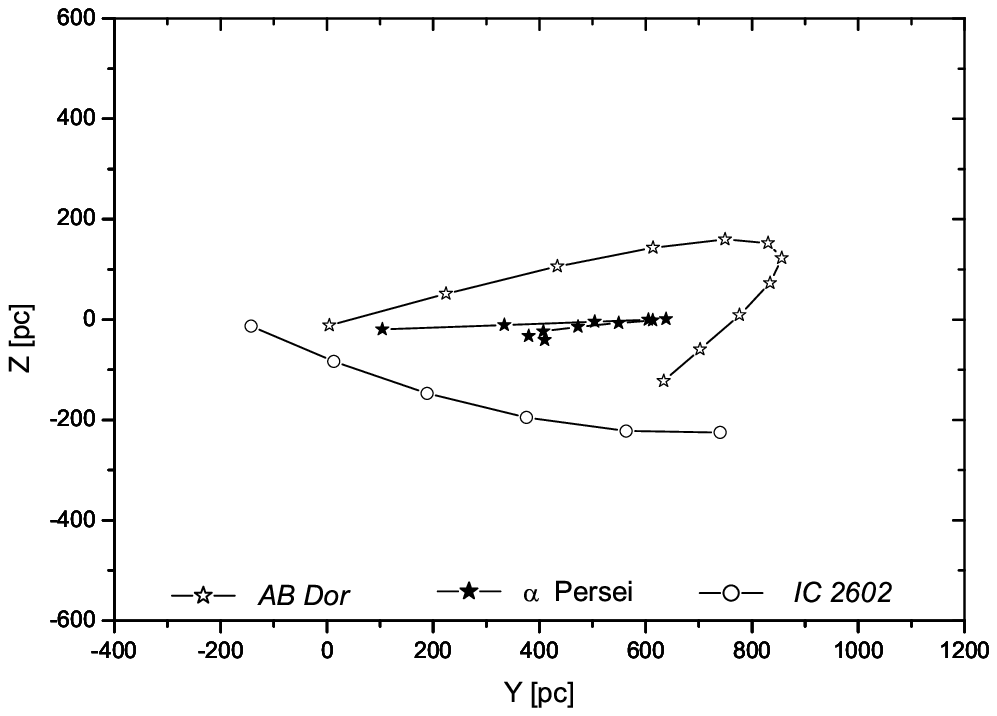} \caption{The same as in Fig 4 in
projection on the YZ Plane. \label{fig4}}
\end{figure}

\begin{figure}
\epsscale{0.7} \plotone{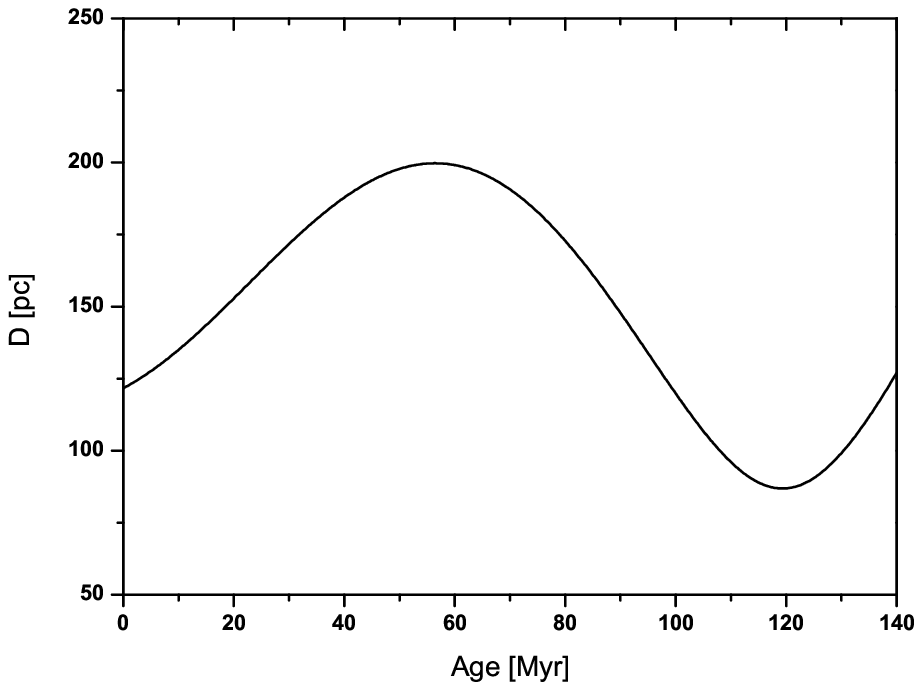} \caption{The evolution of the
distance between the centers of  Pleiades and the AB Dor moving
group. \label{fig5}}
\end{figure}

In the case of clusters it is advisable to work with mean values
of the distances and velocities instead of using the whole sample
of members. This greatly minimizes the errors in the input. In
this work we take the input values for the Pleiades cluster from
Luhman et al. (2005). For the AB Dor group we derive such values
using the Hipparcos data for the members proposed by Zuckerman et
al. (2004a) and more recent published radial velocities. Finally,
the input data for $\alpha$ Persei and IC 2602 were taken from
Robichon et al. (1999).

\section{Dynamical evolution results}

\begin{table}
\centering
  \caption{Relative velocity between the centers of the
Pleiades cluster and the AB Dor moving group for several ages
values near the birth epoch.}
  \begin{tabular}{@{}cc@{}}
    Age   & V [pc Myr $^{-1}$] \\[5pt]
   $-118$   &  $ +0.29 $\\
   $-119$   &  $ +0.06 $\\
   $-120$   &  $ -0.18 $\\
   $-121$   &  $ -0.38 $\\
\end{tabular}
\end{table}

The resulting orbits (Figs. 1, 2, 3, 4) are drawn in the LSR frame
of reference with cartesian axes X, Y, Z positive in the
directions of the Galactic center, the Galactic rotation and the
north Galactic pole respectively. The evolution of AB Dor with the
$\alpha$ Persei and IC 2602 clusters does not attain any minimum.
As seen from Figures 3 and 4 the paths of these systems and that
of AB Dor group are quite different in both projection planes.
Conversely, the calculation for the AB Dor and Pleiades cluster
shows (Figs. 1, 2) that a minimum was reached at the age of $-$119
$\pm$ 20 Myr. In Fig. 5 we show the distance between the centers
of Pleiades and AB Dor as a function of time. In Table 1 we quote
some values of the relative velocity between the centers of these
systems. As can be seen from this table, the relative velocity is
negligible near the estimated epoch (119 Myrs) of Pleiades
formation, showing that here we do not have a simple crossing of
orbits. The minimum separation of AB Dor and the Pleiades cluster
determines a region of 43 $\pm$ 11 pc of radius. As in previous
work (Jilinski et al. 2005) the errors in the age and also in the
radius were estimated through Monte Carlo tests. Tests with 1000
realizations were performed. Because the dynamical age was
computed differentially here, possible systematic errors caused by
uncertainties in the potential will mutually compensate each
other. Hence the resulting errors will be determined by the
uncertainties in the initial data.

The region at minimum is quite large. The size of the region of
maximum orbits approximation is probably connected with an
expansion of AB Dor group during the formation process. The loss
of rotational support of the system resulting from the expansion
is reflected in an increased U velocity, that is, in the velocity
component in the direction of the Galactic center. However, the
orbits of both groups are readily spread out along the positive Y
direction by the action of the Galactic rotation. The groups are
moving towards the Galactic plane along the Z direction .

A scenario for the formation of the Pleiades star cluster
involving the generation of an expanding population has been
inferred by Kroupa et al. (2001) from their N-body calculations.
If this applies also to our case we can propose that the AB Dor
moving group might be the expanding population (group I), or at
least part thereof, present in that scenario. Our results give a
range (100$-$140 Myr) for the age, somewhat different from that
found in Luhman et al. (2005) but not conflicting with their
conclusions concerning the age of the AB Dor group.


\section{Discussion and Conclusions}

\begin{table}
 \centering
 \begin{minipage}{100mm}
  \caption{Individual metallicity determinations for AB Dor stars.}
  \begin{tabular}{@{}lrcl@{}}
 Star        & [Fe/H]&  T$^{0}$K   & Ref  \\[5pt]
 HIP 6276    &  0.06 & 5421 &  Valenti et al. 2005 \\
 CD $-$33 2353&  0.00 & 4550 &  Castilho et al.
2005 \\
 HIP 18859   & -0.11 &  F5V &  Cayrel de Strobel et al. 2001 \\
 HIP 18859   &  0.06 &  F5V &  Cayrel de Strobel et al. 2001 \\
 HIP 18859   & -0.03 &      &  Taylor 2003 \\
 HIP 18859   &  0.06 & 6246 &  Shi et al. 2004 \\
 HIP 18859   & -0.11 & 6162 &  Soubiran et al. 2005 \\
 HIP 18859   & -0.07 & 6308 &  Gray et al. 2003 \\
 CD $-$38 2324 & -0.20 & 5530 &  Cayrel de Strobel et al. 2001 \\
 CD $-$38 2324 &  0.00 & 4800 &  Castilho et al. 2005\\
 HIP 25647   &  0.18 & 5143 &  Cayrel de Strobel et al. 1997 \\
 HIP 26401   &  0.00 & 5400 &  Castilho et al.
2005 \\
 HIP 26373   & -0.02 & 5040 &  Soderblom et al. 1998 \\
 HIP 26373   &  0.00 & 4850 &  Castilho et al. 2005 \\

 HIP 30314   & -0.15 & 5600 &  Castilho et al.
2005\\
 HIP 37855   & -0.10 & 5800 &  Castilho et al.
2005 \\
 HIP 63742   & -0.06 & 5287 &  Gaidos and Gonzales 2002 \\
 HIP 82688   &  0.05 & 5967 &  Valenti et al. 2005 \\
 HIP 107684  &  0.00 & 5500 &  Castilho et al.
2005 \\
 HIP 114530  &  0.00 & 5400 &  Castilho et al.
2005 \\
 HIP 116910  &  0.00 & 5400 &  Castilho et al.
2005 \\
 HIP 118008  & -0.05 & 4850 &  Castilho et al.
2005 \\

\end{tabular}
\end{minipage}
\end{table}

In this work we have considered the question of the age of the AB
Dor moving group from the dynamical point of view. As mentioned in
the introduction, two quite different estimates for the age of
this group have been given: $\sim$ 50 Myr (Zuckerman et al.
2004a), and $\sim$ 125 Myr (Luhman et al. 2005), which is also the
age of the Pleiades open cluster. Here we have approached this
problem by comparing the 3D orbits of the centers of AB Dor, and
of the open clusters Pleiades, $\alpha$ Persei and IC 2602. The
results of this analysis show that the AB Dor moving group and the
Pleiades cluster are coeval having formed some 119 $\pm$ 20 Myr
ago. As for Pleiades, the dynamical age is compatible with the
evolutionary age determinations for this system. For instance, the
upper main sequence turnoff point gives an age of 100 Myr (e.g.
Mermilliod 1981, Meynet et al. 1993) while somewhat more recent
determinations using the lithium boundary method yield an age of
125 Myr (Stauffer et al. 1998, 1999, Barrado y Navascu\'es et al.
1999, 2004). As for the AB Dor group, our results are consistent
with Luhman et al. (2005) who concluded that an age greater than
50 Myr, of the order of the Pleiades age, would help eliminate the
discrepancy between models and measurements of the mass of AB
Doradus C star. It should be remarked, however, that according to
Nielsen et al. (2005) a greater age will be only part of the
solution because of problems related with the effective
temperature of that brown dwarf.

The size of the birth region of Pleiades and AB Dor moving group
 yielded by our dynamical solution could, in principle, be
 explained within the  framework  of the open cluster formation scenario put
 forward by Kroupa et al. (2001). In this scenario a bound and an
 unbound system are born together. We suggest that the unbound system expanding
 at formation could be identified with the AB Dor group.
 This overall picture is also in agreement with conclusions of Luhman et al. (2005) who
 suggested that the AB Dor group could be a remnant of an OB association
 related to the formation of the Pleiades cluster.

Apart from these dynamical characteristics, the Pleiades and AB
Dor group also share some other features that can provide
independent supporting evidence of their common origin. For
example, this appears to be the case with the iron abundance. In
fact, 45 spectroscopic high resolution determinations of the
metallicity (Cayrel de Strobel et al. 2001) give a mean value of
0.00 $\pm$ 0.01 for the Pleiades, whereas for AB Dor a metallicity
of -0.02 $\pm$ 0.02 was obtained as an average from high
resolution spectroscopic measurements in a sample of 15 member
stars, including AB Doradus itself (see Table 2). Comparisons of
lithium abundance are also interesting and may provide additional
support for Pleiades and AB Dor coevality. In this context, da
Silva et al. (2006) have shown that lithium abundances data
indicate a depletion of this element between 11 Myr and the main
sequence. This was done by considering the members of groups
$\beta$ Pictoris, Tucana / Horologium, AB Dor and of the Pleiades
cluster. For all stellar temperatures the lithium abundance of
$\beta$ Pictoris members is larger if compared with those of
Tucana / Horologium, whereas stars of AB Dor and Pleiades have
similar lithium distributions.

 Some possible consequences of the suggested common origin of the
 AB Dor group and the Pleiades cluster deserve special attention.
 Two particulary interesting ones are the stellar mass distribution and the
 fractional distribution of candidates of debris discs in both
 stellar systems. Concerning the mass distribution, it is interesting to
 note the apparent absence of A type stars with masses about 2.5
 M$_{\sun}$ (and also B type stars) in the proposed
 list of members of the AB Dor group by Zuckerman et al. (2004a).
 This in fact, is notorious if we consider that younger
 associations such as TW Hya and $\beta$ Pictoris do contain A
 type stars. If this is not an observational selection effect, we
 can speculate that the absence of A Type stars in the AB Dor group could be
 related to the fact that they are present in Pleiades. An efficient mass
 segregation mechanism should then have been acting during the
 initial separation phase of these groups.

 As for debris discs, a recent 24-micron Spitzer survey (emission
 on these wavelengths is produced in the planetary internal zone between 1 and 20 AU
 from the central star) in the Pleiades by Gorlova et al. (2006)
 has shown that 25\% of B-A type members and 10\% of F$-$F3 type stars
 are debris disk candidates due to their excess IR emission at an age
 around 100 Myr. It would then be interesting to conduct a similar survey at 24
 micron in the AB Dor group and compare its fraction of
 debris disc candidates with that in Pleiades.

\acknowledgments

EJ thanks MCT of Brazil for the financial support under the
contract 384222/2006$-$4.

\end{document}